\documentstyle[referee]{laa}
\begin{document}
 \input epsf
\newcommand{\dix}{\,{ \,10} }
 \newcommand{\mic}{\,{ \mu m} }
 \newcommand{\pac}{\,{ pc} }
 \newcommand{\Wmsr}{\,{ Wm^{-2}sr^{-1}} }
 \newcommand{\Wmsrmic}{\,{ Wm^{-2}sr^{-1}\mic^{-1}} }
  \newcommand{\Wmsrhz}{\,{ Wm^{-2}sr^{-1}Hz^{-1}} }
 \newcommand{\Wcmsr}{\,{ Wcm^{-2}sr^{-1}} }
 \newcommand{\Wcm}{\,{ Wcm^{-2}} }
  \newcommand{\cm}{\,{ cm^{-2}} }
  \newcommand{\cmmun}{\,{ cm^{-1}} }
 \newcommand{\Wm}{\,{ Wm^{-2}} }
 \newcommand{\WH}{\,{ W/H_{atom}} }
 \newcommand{\Wcmmic}{\,{ Wcm^{-2}\mic^{-1}} }
 \newcommand{\Wmmic}{\,{ W/m^2/\mic} }
 \newcommand{\MJysr}{\,{ MJysr^{-1}} } 
 \newcommand{\Wcmmicsr}{\,{ W/cm^2/\mic /sr} }
 \newcommand{\Dl}{\,{ \Delta \lamb} }
 \newcommand{\lamb}{\,{ \lambda} }
\newcommand{\Inu}{\,{ I_{\nu}}}
 \newcommand{\nuInu}{\,{ \nu I_{\nu}}}
 \newcommand{\nHdeux}{\,{n_{H_2}} }
\newcommand{\nHtwo}{\,{n_{H_2}} }
\newcommand{\NHdeux}{\,{N_{H_2}} }
\newcommand{\nHII}{\,{n_{HII}} }
\newcommand{\nH}{\,{n_{H}} }
\newcommand{\NH}{\,{N_{H}} }
\newcommand{\cmcube}{\,{ cm^{-3}} }
\newcommand{\cmdeux}{\,{ cm^{-2}} }
\newcommand{\cmdeuxkms}{\,{ cm^{-2}(km/s)^{-1}} }
\newcommand{\cmun}{\,{ cm^{-1}} }
\newcommand{\cmcubes}{\,{ cm^{-3}s^{-1}} }
\newcommand{\kcmcubes}{\,{ cm^{3}s^{-1}} }
\newcommand{\tUV}{\,{\tau_{\mbox{UV}}} }
\newcommand{\tV}{\,{\tau_{\mbox{V}}} }
\newcommand{\Lsol}{\,{ L_{\sun}} }
\newcommand{\IO}{\,{ I_0} }
\newcommand{\IL}{\,{ I_L} }
\newcommand{\epstrois}{\, {\varepsilon_{3.3}} }
\newcommand{\Hbeta}{\,{ H_{\beta}} }
\newcommand{\Halpha}{\,{ H_{\alpha}} }
\newcommand{\Bralpha}{\,{ Br_{\alpha}} }
\newcommand{\pccmsix}{\,{ pc \, cm^{-6}} }
\newcommand{\Htwo}{\,{ H_2} }
\newcommand{\CeighteenO}{\,{C^{18}O} }
\newcommand{\Av}{\,{ A_V} }
\newcommand{\Cplus}{\,{ C^+} }
\newcommand{\Etrois}{\, {E_{3.3}} }
\newcommand{\FC}{\, {F_{\lambda}(C)} }
\newcommand{\FK}{\, {F_{\lambda}(K)} }
\newcommand{\FL}{\, {F_{\lambda}(L)} }
\newcommand{\FPAH}{\, {F_{\lambda}(PAH)} }
\newcommand{\hnu}{\, { h\nu}}
\newcommand{\douzeCO}{\, { ^{12}CO}}
\newcommand{\treizeCO}{\, { ^{13}CO}}
\newcommand{\Hplus}{\, { H^+}}
\newcommand{\HdeuxO}{\,{ H_{2}0}}
\newcommand{\NHtrois}{\,{ NH_{3}}}
\newcommand{\orthoNHtrois}{\,{ ortho-NH_{3}}}
\newcommand{\paraNHtrois}{\,{ para-NH_{3}}}
\newcommand{\HtwoeighteenO}{\,{ H_{2}^{18}O}}
\newcommand{\HtwoO}{\,{ H_{2}O}}
\newcommand{\orthoHtwoO}{\,{ortho-H_{2}O}}
\newcommand{\paraHtwoO}{\,{para-H_{2}O}}
\newcommand{\Odeux}{\,{ O_{2}}}
\newcommand{\Otwo}{\,{ O_{2}}}
\newcommand{\Fd}{\,{ F_d}} 
\newcommand{\betal}{\,{ \beta_{l}}}
\newcommand{\betad}{\,{ \beta_{d}}}
\newcommand{\gu}{\,{ g_{u}}}
\newcommand{\gl}{\,{ g_{l}}}
\newcommand{\Eu}{\,{ E_{u}}}
\newcommand{\El}{\,{ E_{l}}}
\newcommand{\Aul}{\,{ A_{ul}}}
\newcommand{\nup}{\,{ n_{u}}}
\newcommand{\nl}{\,{ n_{l}}}
\newcommand{\smun}{\,{ s^{-1}}}
\newcommand{\Ho}{\,{H_0}}
\newcommand{\apj}{\,{ApJ}}
\newcommand{\Cl}{\,{C_l}}
\newcommand{\degres}{\,{ ^{\scriptsize{o}}}}

    \thesaurus{Sect. 22         
              (   )
             }
   \title{All sky mapping of the Cosmic Microwave Background at $8\arcmin$ 
angular resolution with a 0.1 K bolometer: simulations}
 
   \subtitle{}

   \author{M. Giard \inst{1}, E. Hivon \inst{2,4}, C. Nguyen \inst{1}, 
   R. Gispert \inst{3}, K.M. G\'orski \inst{2,5}, A. Lange \inst{4}, 
   and I. Ristorcelli \inst{1}}

   \offprints{M. Giard}
 
   \institute{
 Centre d'Etude Spatiale des Rayonnements, 9 avenue du Colonel Roche, BP
4346, F-31028 Toulouse Cedex 04, France.
\and Theoretical Astrophysics Center, Copenhagen, Denmark.
\and Institut d'Astrophysique Spatiale, Orsay, France.
\and Department of Physics, Math, and Astronomy, California Institute of Technology, USA.
\and Warsaw University
Observatory, Warsaw, Poland.
}
 
   \date{Received               ; accepted              }
 
   \maketitle
 
   \begin{abstract}  We present simulations of observations 
   with the 143 GHz channel of the 
   Planck High Frequency Instrument (HFI). These simulations
are performed over the entire sky, using
   the true angular resolution of this channel:  
   8 arcmin FWHM, 3.5 arcmin per pixel.  
   We show that with measured 0.1 K bolometer performances
   , the sensitivity needed  on the Cosmic Microwave Background (CMB) survey is obtained using 
  simple and robust data processing techniques, including a destriping
algorithm.


   \end{abstract}
 
%
\section{\label{introduction} Introduction}

The ESA Planck mission, to be launched in 2007, will
perform a quasi complete survey of
the Cosmic Microwave Background (CMB)
with an angular resolution reaching $5 \arcmin$.
It will use a 1.5 metre diametre off-axis telescope
placed at the lagrangian L2 point, about 1.5 million
kilometres from the Earth, in the antisolar direction.
The focal plane includes two instruments, LFI and HFI,
using respectively HEMTS and bolometer technologies. 
The frequency range is from 30 to 100 GHz  for
the LFI
(beam sizes from $30 \arcmin$  to $10 \arcmin$), 
and from 100 to 857 GHz for the HFI ($11 \arcmin$ to $5 \arcmin$) .     
(for a detailed description see Bersanelli et al. \cite{Bersanelli96} 
and the Web pages at
http://astro.estec.esa.nl/planck/). 
Planck will allow to grasp details on the last
scattering surface which are more
than ten times smaller than the size of the horizon at the epoch of 
recombination. The observed pattern will inform us on the statistics of the primordial
seeds that have been generated in the very early universe, providing a unique observational
test to fundamental physics. In addition, the maps will probe
the Universe to an unprecedented distance, $z \simeq 1000$,
allowing to derive the important cosmological parameters
which determine its geometry (see e.g. Bond et al. \cite{Bond97} and
White et al. \cite{White98}). 

There are many observational difficulties to this challenge.
Some are astrophysical (contamination by foregrounds) and others
are instrumental  (e.g. far sidelobe pickup and detector noises). The effects
of the astrophysical foregrounds have been discussed in various papers
and it has been demonstrated that the CMB can be recovered down to the
level of the cosmic variance if adequate frequency bands
are used (see e.g. Gispert and Bouchet \cite{Gispert96}, or Linden-Vornle and Norgaard-Nielsen \cite{Linden98}).
The instrumental limitations seem to be a much more uncertain issue.
The detector behaviour is one of the most critical instrumental issues. 
Janssen et al. \cite{Janssen96} investigated the problem of 1/f noise 
contamination for the
PSI and FIRE proposals to NASA's Medium-class Explorer
program. Their study demonstrated that, given the performances of the 
detectors (bolometers and HEMTs), a simple scan by scan drift removal was
enough to subtract most of the temporally correlated noise (e.g. 1/f noise)
without any significant alteration of the CMB power spectrum. A more sophisticated
method presented by Delabrouille (\cite{Delabrouille98b}) uses the scans intersects
redundancy to determine and subtract the low frequency useless signals. Except
in very recent works (see the draft by Maino et al. \cite{Maino99}), up
to now, mostly for computational reasons, this destriping has been simulated
at a very limited angular resolution (about 1 degree).

In this paper we present all sky simulations of CMB observations 
in the 143 GHz channel of the HFI instrument 
with a 0.1K bolometer. This channel is simulated with its true
beam width, 8 arcmin full width at half maximum (FWHM), using sky maps with a
pixel size of 3.5 arcmin. We show that a simple data processing (pixels
binning) 
including a destriping algorithm, allows to efficiently eliminate the 1/f noise. 
The next section shows the measured performances
of Planck prototype bolometers mounted on a 0.1 K open cycle
 $^3He/^4He$ dilution refrigerator. Section \ref{simulations} deals
 with the simulations of all-sky Planck observations and map reconstruction. 
The results are discussed  in section \ref{results}.

\section{\label{bolometers} Performances of 0.1K bolometers}

Prototype spiderweb bolometers with time
constants suitable for Planck, e.g. a few milliseconds, have been integrated
on a 0.1 K open cycle dilution refrigerator. The readout system 
uses a square-wave current control of the bolometer thermistor with
capacitive load. This full digitally controled system is described in 
Gaertner et al. \cite{Gaertner97}.
It has been optimized to mimize power consumption, given the Planck needs.
Those are: time frequency coverage from 0.01 to 200 Hz and 
sensitivity of order $3 nV/Hz^{0.5}$ (corresponding to the typical 
Johnson noise of the bolometer thermistor, a few megohms).
Fig. \ref{fig1} shows the electrical noise performance of this system
tested on a 300K resistor (a), and on a 0.1K Planck prototype 
bolometer (b). The measurement on the resistor demonstrates
that the performance of the modulation control system plus
amplifier chain fullfills the noise level requirement. 
The bolometer noise power spectrum has been obtained with an active temperature
control of the 0.1 K plate. Volts have been calibrated against watts  using
the bolometer electrical response derived from the I(V) curve:

\begin{figure}[ht]
\vbox to 10cm{
\epsfxsize=14cm
\epsfbox{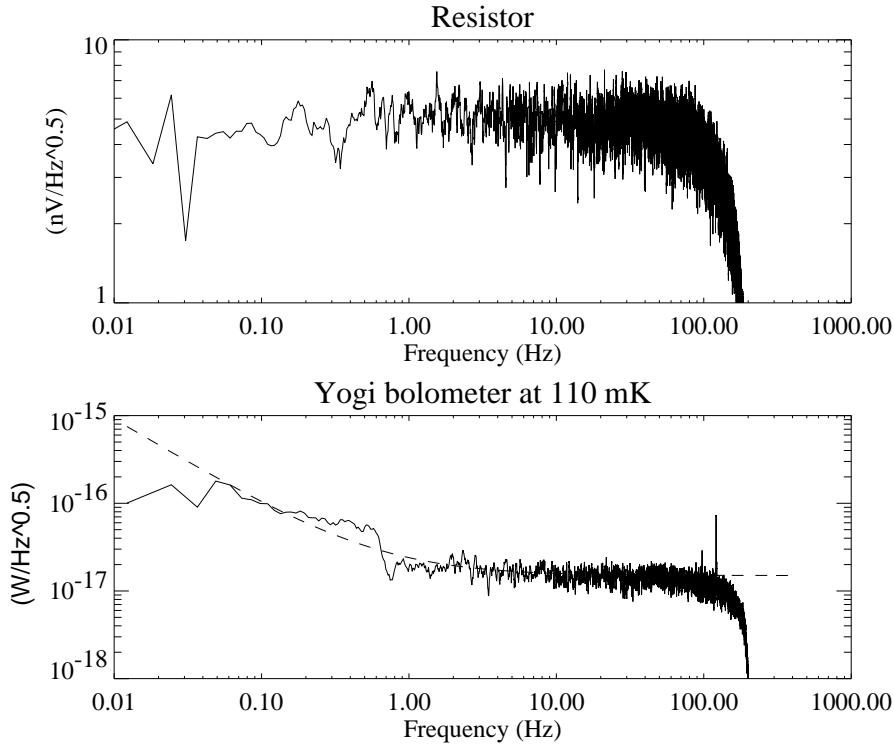}
}\caption[]{Measured noise performances on a 0.1 K prototype spiderweb
bolometer (lower), and a pure resistor (upper). 
The 1/f noise on the bolometer is due to temperature unstabilities
of the 0.1 K stage. The dashed line on the
lower plot is the noise power spectrum used in the simulation, $f_{knee} = 0.6 Hz$
(see text).  
\label{fig1}}
\end{figure}

\begin{eqnarray}
\frac{dV}{dP} (volt/watt) = \frac{R-Z}{2RI}
\label{eq1}
\end{eqnarray}

where  $Z = dV/dI$ is the dynamical impedance of the bolometer and $R = V/I$
is its resistance.
The bolometer sensitivity in this system is of the order of $\dix^{-17}
W/Hz^{-0.5}$ for all frequencies larger than 1 Hz. The excess noise at
lower frequencies has been attributed to residual temperature fluctuations
of the 0.1 K system. Its thermal origin is demonstrated by the correlation
with the bolometer response: this noise is always at the same level in
terms of "Watts" for different system temperatures, thus bolometer responses,
whereas the high frequency plateau in the noise spectrum has a constant level
in terms of "volts", whatever is the bolometer response.
Most of the 0.1 K temperature fluctuations are due to
instabilities in the flow of the cooling fluids. 
Current studies of the 0.1K refrigerator system aim at 
reducing this problem together with improving the
active temperature control of the bolometer plate.
However, our simulations in 
section \ref{simulations} show that this noise level is acceptable and 
already allows to measure the CMB signal with the required accuracy.

\begin{figure}[ht]
\vbox to 10cm{
\epsfxsize=14cm
\epsfbox{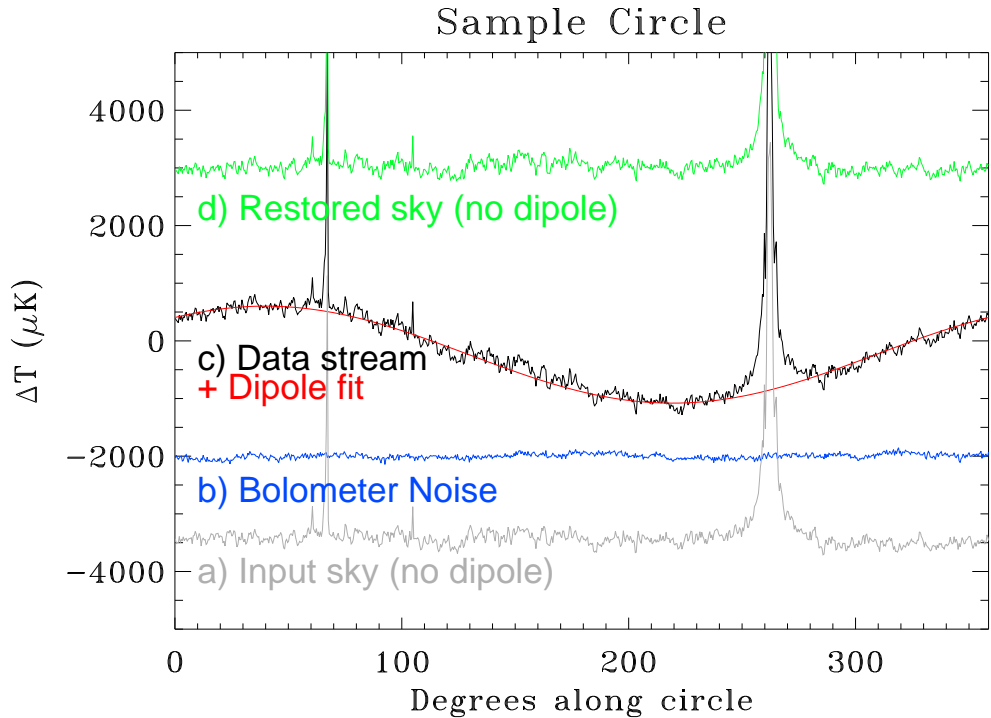}
}\caption[]{A sample simulated circle showing the co-added signal
from a single 143 GHz bolometer between two satellite pointings.
\label{fig2}} 
\end{figure}

\section{\label{simulations} All sky simulations}

\subsection{\label{datastream} Simulating the data stream} 
 
We have simulated 12 months of the Planck mission with the following
observing strategy:  the satellite spin rate is $S=1$ rpm, 
with the spin axis following the 
antisolar direction. The beam scans the sky at an angular distance 
of 85 degrees from the spin axis. Re-pointing of the spin axis
to follow the antisolar direction is
performed with a step of 2.5 arcmin (i.e. a period  $T \simeq 61 minutes$).
The spin axis follows a sinuso\"{\i}dal
trajectory along the ecliptic with full amplitude $\Delta\delta = \pm 10 $ degrees,  
so that the polar caps are not left unobserved. A duration of 12 months
is the minimum which allows to observe twice at 6 months
interval almost all directions of the sky, thus providing redundancy.

In this simulation, the observed sky is the sum of the CMB, including the COBE-DMR determination of the CMB dipole
(Lineweaver et al. \cite{Lineweaver96}), and the Galaxy extrapolated from the composite 100 $\mic$
IRAS-COBE/DIRBE all sky dataset (Schlegel et al. \cite{Schlegel98}). This extrapolation
assumes a dust
spectrum typical
of high latitude cirrus clouds, $T_d = 17K, n_d = 2$ (Boulanger et al. \cite{Boulanger96}). 
The CMB sky is obtained from a standard Cold Dark Matter model:
$\Omega_{tot} = 1$, $\Omega_b = 0.05$, $\Omega_{\Lambda} = 0$,
 and $ \Ho = 50 km/s/Mpc$.
We compute the $\Cl$ power spectrum with CMBFAST
(Seljak et al. \cite{Seljak96}) and the map is generated in a Healpix-type 
all-sky pixellisation
with a pixel size of 3.5 arcmin (G\'orski  \cite{Gorski98}, see
the web pages at http://www.tac.dk/\~{}healpix).
Since our simulations will be limited to the frequencies
of the HFI instrument, $\nu \ge 100 GHz$, we do not
include the galactic synchrotron and free-free emissions.
The COBE/DMR data analysis from Kogut et al. (\cite{Kogut96}) 
has shown that both are smaller than the dust at such frequencies, 
the former being highly correlated with the dust.
The radio point sources are not included either, and they are
supposed to be extracted on a scan by scan analysis. 
The instrumental noise is added after pickup of the sky signal on  a scan
by scan basis. The noise power spectrum follows the characteristics
of our 0.1 K bolometer system, and is the sum of a white noise and 1/f
low frequency noise:

\begin{eqnarray}
N = N_\infty*(1+f_{knee}/f)
\label{eq2}
\end{eqnarray}

With typical values being, $N_\infty = 52 \mu K/Hz^{0.5}$ (physical temperature) 
for the millimeter channels, and $f_{knee} = 0.6 Hz$.

\begin{figure}[ht]
\vbox to 10cm{
\epsfxsize=14cm
\epsfbox{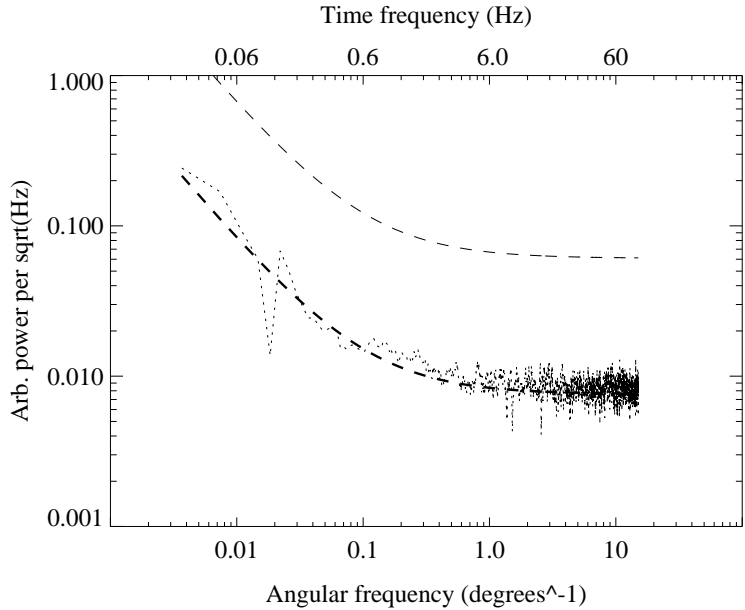}
}\caption[]{The power spectrum of a simulated circle obtained by averaging
64 scans (dots) is compared to the average power spectrum of a single scan (dashed
line) and the average scan power spectrum divided
by $\sqrt 64$ (thick dashed line). For FFT efficiency 64 was used instead of
60, the exact number of scans per circle.
\label{fig7}} 
\end{figure}

In order to spare computer time, only one scan 
(called circle in the following) is simulated between two
satellite re-pointings. The noise level is correspondingly reduced by  factor
$(T/S)^{0.5} = 7.8 $, assuming that all the scans are coadded into
a single circle and that  
the noise is not correlated from one scan
to the other. This is not strictly correct given the frequency aliasing 
produced by the co-addition (see Janssen et al. \cite{Janssen96}). 
However we have checked that this is correct within the statistics of the noise. 
Fig. \ref{fig7} demonstrates this: we have plotted here the power
spectrum of the circle obtained by the averaging of 64 scans (one hour
of observing time) and compare it to the average power spectrum of a single scan
(i.e. Equ. \ref{eq2}) divided by the square root of the number of scans
coadded.
The data sampling along the scan direction is equal to 
the depointing step: 2.5 arcmin. The time frequency domain is thus sampled
from 0.016 Hz (spin frequency) to 144 Hz (sampling rate).

\subsection{\label{dataproc} Data processing and destriping algorithm}

To construct the simulated sky maps, the data from the circles are simply coadded
within the pixels of the map. As usual with bolometric devices, the absolute level
of the measured power is unknown. 

At zero order, the level of each circle is adjusted
by subtraction of a constant so that the measured signal fits a cosecant law
(galactic dust) plus dipole emission:

\begin{eqnarray}
M_{ik} = r_k (Dipole_{ik} + \frac{g_k}{sin(bII_i)} + C^0_k)
\label{eq3}
\end{eqnarray}

k is the circle index, and i the data index along the circle. 
The free parameters are: $g_k$
the level of the galactic emission for this circle, and $C^0_k$ the constant
to be subtracted. The detector response, $r_k$, is supposed to be fixed and known from 
calibration procedures. Portions of the circle with a strong signal ( $>  6 mK$
or $|lII| < 10^\circ$ )
or a strong gradient (point sources) are excluded from the fit.
 The cosecant law
is introduced in the fit to avoid subtraction of the average galactic signal,
which is positive.
The calibrated data projected on the final survey map is then:

\begin{eqnarray}
D_{ik} = M_{ik}/r_k - C^0_k - Dipole_{ik}
\label{eq4}
\end{eqnarray}

Fig. \ref{fig2} shows a sample simulated circle (addition of about 60 scans)
for one bolometer of the frequency band centered at 143 GHz. 
The dipole and the Galaxy
are the dominating signals. On this circle the Galactic
plane is intercepted at Galactic longitudes of about 18 and 195 degrees.
The primary fluctuations of the CMB are clearly visible on the
measured signal for this single bolometer measurement.

This coarse evaluation of the offset of each circle  is not sufficient
to obtain the high quality maps required for the Planck mission.
We further destripe the data using an algorithm derived from 
Delabrouille (\cite{Delabrouille98b}) which uses the scan intercepts.
We actually adjust the constants $C^1_k$
to subtract to each circle $k$ by minimising the spread between the
measurements contributing to the same sky pixel. This is done
globally for all the sky pixels using for each pixel a weight
wich is proportional to the number of circles contributing to
this pixel. The quantity to be minimised thus reads:

\begin{eqnarray}
J = \sum_{ipix \, \in \, Sky} \left[  \frac{\sum_{ik \, \in \, ipix} (D_{ik} - C^1_k)^2}{n(ipix)} -
 \left(  \frac{\sum_{ik \, \in \, ipix} (D_{ik} - C^1_k)}{n(ipix)} \right)^2 \right]
\label{eq5}
\end{eqnarray}

$ik \, \in \, ipix$ means that the sum is over all measurements
from circle k and index i contributing to pixel ipix.

$n(ipix)$ is the total number of measurements contributing to a
given sky pixel.

We find the minimum of $J$ using a maximum gradient iterative algorithm. At each iteration
the optimal step length in the direction of the gradient can
be computed exactly because $J$ is a quadratic form. The gradient reads:

\begin{eqnarray}
\frac{\partial J}{\partial C^1_k} = 2 \sum_{i \, \in \, k} \left[ 
\frac{\sum_{i'k' \, \in \, ipix(ik)} (D_{i'k'} - C^{1}_{k'})}{n(ipix(ik))} 
- (D_{ik} - C^1_k)
\right]/n(ipix(ik))
\label{eq6}
\end{eqnarray}

$i \, \in \, k$ means that the sum is over all data points from circle k, and 
$i'k' \, \in \, ipix(ik)$ means that the sum is over all measurements
falling on the same sky pixel ipix(ik) as measurement ik.

The optimal iterative step is:

\begin{eqnarray}
\Delta C^1_k = \alpha \frac{\partial J}{\partial C^1_k} 
\end{eqnarray}

with

\begin{eqnarray}
\alpha = \frac{\sum_{ipix \, \in \, Sky} \left[ \frac{\sum_{ik\,\in\,ipix} \frac{\partial J}{\partial
        C^1_k}}{n(ipix)}
    \frac{\sum_{ik\,\in\,ipix} (D_{ik}-C^1_k)}{n(ipix)} - \frac{\sum_{ik\,\in\,ipix}
      \frac{\partial J}{\partial C^1_k}
      (D_{ik}-C^1_k)}{n(ipix)}\right]} 
{\sum_{ipix \, \in \, Sky} \left[ \left( \frac{\sum_{ik\,\in\,ipix} \frac{\partial J}{\partial C^1_k}}{n(ipix)}\right)^2 -
  \frac{\sum_{ik\,\in\,ipix} \frac{\partial J}{\partial C^1_k}^2}{n(ipix)}  \right]}
\label{eq7}
\end{eqnarray}

\begin{figure}[th]
\vbox to 10cm{
\epsfxsize=14.cm
\epsfbox{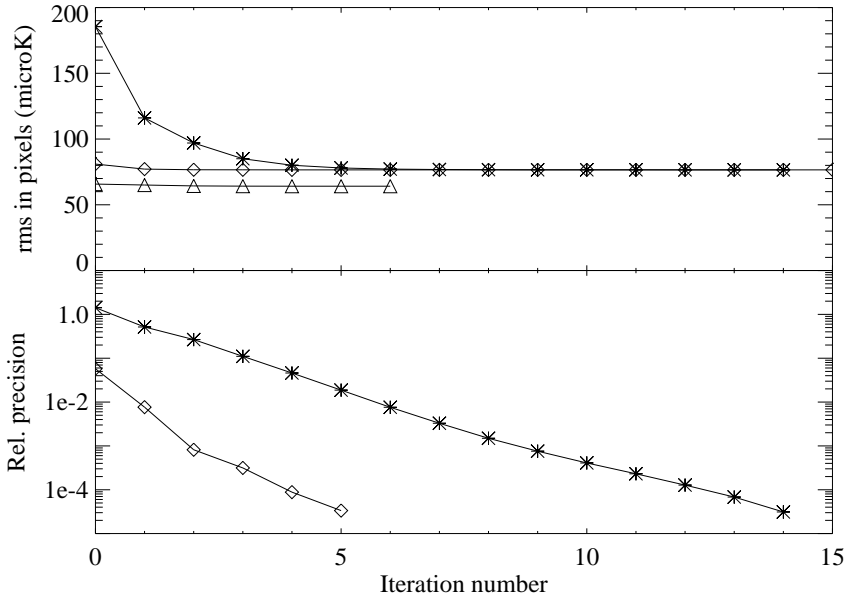}
}\caption[]{Convergence of the iterative maximum gradient algorithm
used to destripe the data. The upper plot shows the average rms spread of the
data points falling into a single pixel as a function of the iteration
step. The stars show the convergence from a starting point where 
each circle has been set to a null average.
The diamonds are obtained with a starting point where the constant
subtracted to each circle is evaluated in a fit which
includes the CMB dipole plus a cosecant law (see equ.
\ref{eq4}). The triangles are obtained for a pure white bolometer noise.
The lower plot shows the relative difference between
the current and final rms value. 
\label{fig8}} 
\end{figure}

Figure \ref{fig8} shows the convergence of our iterative destriping method 
which evaluates
the best constants to subtract to each circle. We plot in this Figure the average over the
whole sky of the rms spread of the data within a pixel (i.e. $sqrt(J/npix)$).
This is shown for two
different starting conditions: i) the stars are obtained with a simplistic
starting condition which consists in forcing the average signal of each circle to be nul,
ii) the diamonds are obtained starting from circles where the constant
subtracted has been set using (\ref{eq3}): a fit which includes the CMB dipole and a
cosecant galactic law. This second method is very efficient and a relative precision 
better than 1.e-4 is obtained within five or six iterations of the
starting point. The value of 1.e-4 for the relative precision on
the final average rms is the criterium that we have fixed to end the
destriping iterations. The triangles show the result for a pure white bolometer
noise (no 1/f noise). The comparison of the final rms values for both
cases shows an excess of about 15\% for the 1/f noise with 
respect to the white noise. This is to be compared to the noise excess induced
by the 1/f tail : 10\% on the rms for a knee frequency of 0.6 Hz if we take
into account only the frequencies higher than the spin rate (0.016 Hz).
In fact the noise at frequencies smaller than the spin rate has not been
fully washed out since we only subtract a constant level to each circle.
A Wierner filtering in the frequency domain would be optimum, but 
this would alter the astrophysical signal. 

Our simulations are performed on a Pentium III 500 MHz processor 
using the IDL environment. The data are first generated for the full mission
and stored in a file with the corresponding sky pixel indexes. This needs
about 30 minutes of computer time. Then the iterations are performed, which needs about 20 
minutes each, so that the time for a full simulation with destriping is less
than 2 hours. 

The computation of the 
spherical harmonics power spectrum from the simulted map is about
the same duration, using the fast code implemented with the Healpix package
(Hivon and G\'orski \cite{Hivon98}).
The signal at galactic latitudes bellow 30 degrees is set to zero for the 
$\Cl$ power spectrum estimation. This implies some aliasing of the
galactic component for the low l values (see below).

\begin{figure}[th]
\vbox to 10cm{
\epsfxsize=14.cm
\epsfbox{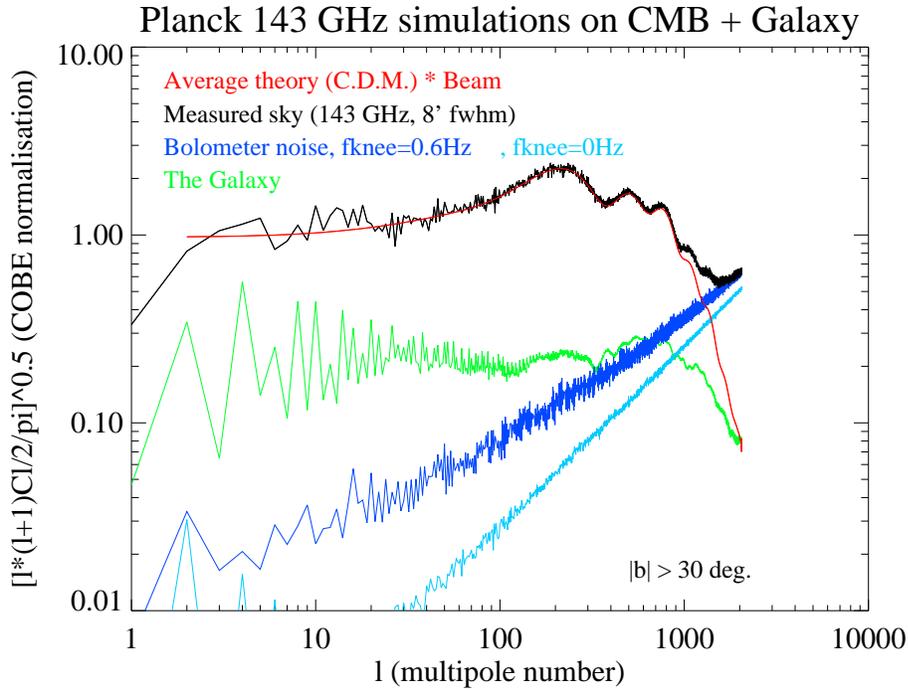}
}\caption[]{Black: $\Cl$ transform of the simulated reconstructed sky at 143 GHz
for 6 bolometers, 12 months (galactic latitudes bellow 30 degrees have been set
to zero). Red: average theory from which the CMB sky was generated
(convolved by the instrumental beam: 8 arcmin FWHM gaussian). 
Green: contribution from the galactic dust.
Blue: contribution from the bolometer noise with 1/f noise (dark)
and for pure white noise (light).
\label{fig5}} 
\end{figure}

\section{\label{results} Results and conclusion}

Fig. \ref{fig3}  shows the allsky average map for six bolometers 
at 143 GHz. It has been obtained by averaging six simulations
with independent noise realisations. This is certainly a simplistic 
assumption since one can expect the bolometer noises 
to be correlated. However, such correlated noises will be monitored
by thermometers and blind bolometers.
The average signal-to-noise ratio on each 3.5 arcmin pixel of
this map is about 6 at high galactic latitudes, 
where the CMB dominates, the rms signal and noise being respectively
$110$ and $22 \mu K$.  The variance of the reconstructed map for
a single bolometer has been computed using a Monte-Carlo method
which repeats the simulation for different noise seeds.
Fig. \ref{fig4a} and \ref{fig4b} show respectively the 
variance for the non-destriped 
 and destriped map estimated from 28 simulations. 
The gain induced by the use of the destriping 
is clearly visible. 

The noise due to the data processing has been
estimated in a simulation performed with no instrumental noise. 
The destriping has been iterated 6 times, which is the average number
of iterations needed to reach the convergence criterium in the case
where the instrumental noise is included. The
map obtained by difference from this simulation and the input
sky is shown in Fig. \ref{fig4c}. The level of this difference
is always below a few $\mu K$, demonstrating that the data
processing noise is negligible compared to the instrumental noise.
Moreover, its level can be lowered by several orders of magnitude by simply
increasing the number of iterations in the destriping algorithm.

We have also computed the noise correlation of a
particuliar pixel with all other sky pixels. The pixel is located at
galactic coordinates (LII, BII) = ($45\degres$, $28\degres$). The 
correlation coefficient  of this pixel with others is shown for the sky region
arround that pixel in Fig. \ref{fig10} a and b (respectively
non-destriped and destriped). The number of iterations for this estimation
has been set to 128.
It needs to be much higher than for the auto-covariance since the average value
is in general close to zero. The correlation of the central pixel with its
neighbours is, as expected, maximum along the circles intercepting this pixel.
It is clearly improved with the iterative destriping. The spread in the
values of the estimated correlation for the other pixels is actually 
dominated by the error of the Monte-Carlo method : about $5 \dix^{-10} K^2$
($rms^2$) for 128 iterations and a rms noise in this part
of the sky of $70 \dix^{-6}$ K.

\begin{figure}[th]
\vbox to 10cm{
\epsfxsize=14.cm
\epsfbox{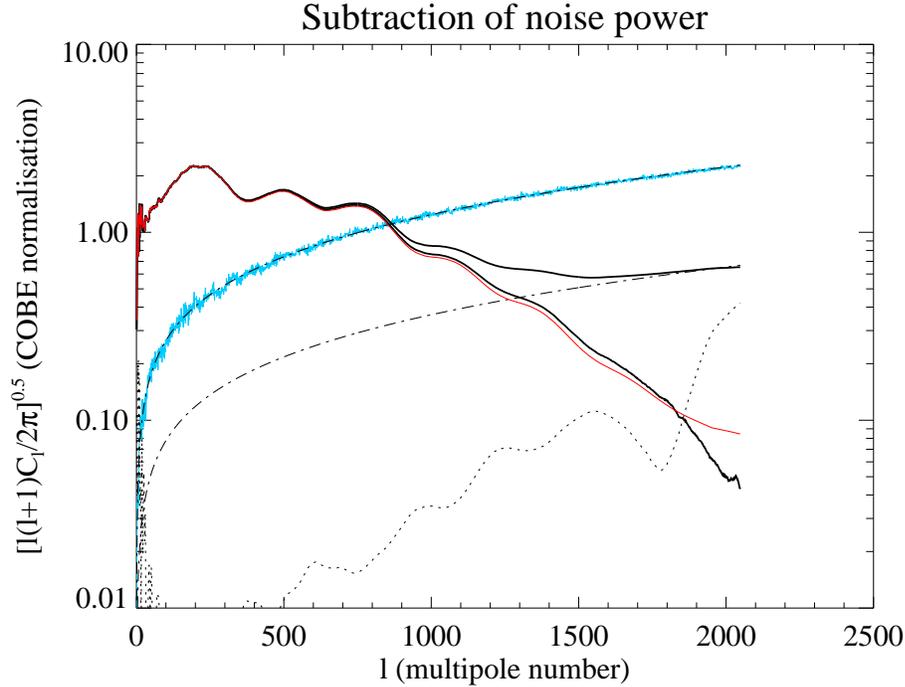}
}\caption[]{Recovered smoothed $\Cl$ for the 12 months simulation (thick black) compared to the original $\Cl$ (red) at 143 GHz. Blue: pure noise from a 2 bolometer difference fitted 
by a log-log 3rd order polynomial (black dashed) and adjusted to the measured
$\Cl$ at high l's (black dot-dashed). Black dots : relative error on the
recovered $\Cl$.
\label{fig6}} 
\end{figure}

The $C_l$  power spectra of the reconstructed sky and the
various components are plotted in Figure \ref{fig5}.
We did not try here to correct for the beam of the instrument. 
Our simulated beam is a gaussian of 8 arcmin FWHM.
This could be removed in $\Cl$ space by division of 
the measured spectrum by the $\Cl$ transform of the
beam. However, the true beam will not be
symmetrical (due to the off-axis telescope and the electrical signal
filtering in the scanning direction) so that such a 
method will provide only a first order approximation.
One will actually need more sophisticated inversion methods
which are out of the scope of this paper.

The bolometer noise dominates the power-spectrum at high multipole numbers.
Its estimation and subtraction will be a critical step for the
recovery of the astrophysical sky power-spectrum for these
multipole numbers. The shape of the noise power spectrum can practically
be estimated from a pure noise map obtained by difference from 
2 bolometers. Its level can then be adjusted to the 
measured (sky+noise) power-spectrum at high multipole numbers
and subtracted. Fig. \ref{fig6} compares the recovered power
spectrum to the original one. The noise power spectrum from a 2 
bolometer difference has been
fitted by a 3rd order polynomial in log-log scale and then quadratically
subtracted to the sky measured power-spectrum, 
after adjustment on high multipole
numbers, $l > 1950$. In this Figure the final and original $\Cl$'s have
been smoothed by a square window of constant logarithmic width (dl/l = 0.1),
thus reducing the fluctuations due to the cosmic variance. 
The relative error on the recovered $\Cl$ is also shown in Fig. \ref{fig6}.
Its level is below 1 \% for $l < 500$ , and rises from 1 \% to 10 \% for 
$500 < l < 1950$. 

The galactic emission,
which has not been subtracted in our data processing, dominates 
the errors at high multipole numbers.
Informations from the Planck higher frequency channels will
be used to subtract this component. The accuracy of this
subtraction will be limited by the uncertainty in the values of the 
dust spectral index and temperature deduced
from the submillimeter channels. 
A conservative estimate of the residual galactic
signal can be obtained if we assume that the relative
error on the estimation of the galactic signal
is of the order of 10 \%.
 With this hypothesis the galactic residual
will never be higher than 3 \% of the CMB
up to the higher l values ( $l < 2048$, see Fig. \ref{fig6}). 

The simulations presented in this paper 
address a limited number of points,
both with regard to the
astrophysical sky (we include only the CMB and the galactic
dust which are dominant at 143 GHz) and to the instrument 
(only one frequency channel instead
of the 9 channels of the two Planck focal plane instruments).
As mentioned in Sect. \ref{introduction} there
will be systematic signals produced by temperature 
variations in the payload.
The galactic signal picked up in the far side lobes
of the telescope will be significant, even at high
galactic latitudes. A proper analysis of the data,
with optimal rejection of the systematics and noise sources
for reconstructing the sky maps, will have to involve
a global strategy making use of all frequency channels. 
Our result, however, show that even a single channel
analysis at 143 GHz, which assumes noise performances
measured on a prototype bolometer mounted on a 0.1 K open cycle
dilution refrigerator, produces satisfactory results.

\acknowledgements{
Map projections routines and power spectrum calculations were performed with
the Healpix package written by G\'orski, Hivon and Wandelt
(http://www.tac.dk/~healpix).
KG and EH were funded by the Dansk Grundforskningsfond through its
funding for TAC.
E.H. and A.L. were supported in part by NASA Grant NAG5-6573.}

\bibliographystyle{astron}

\begin{figure}[ht]
\vbox to 20cm{
\epsfxsize=13.5cm
\epsfbox{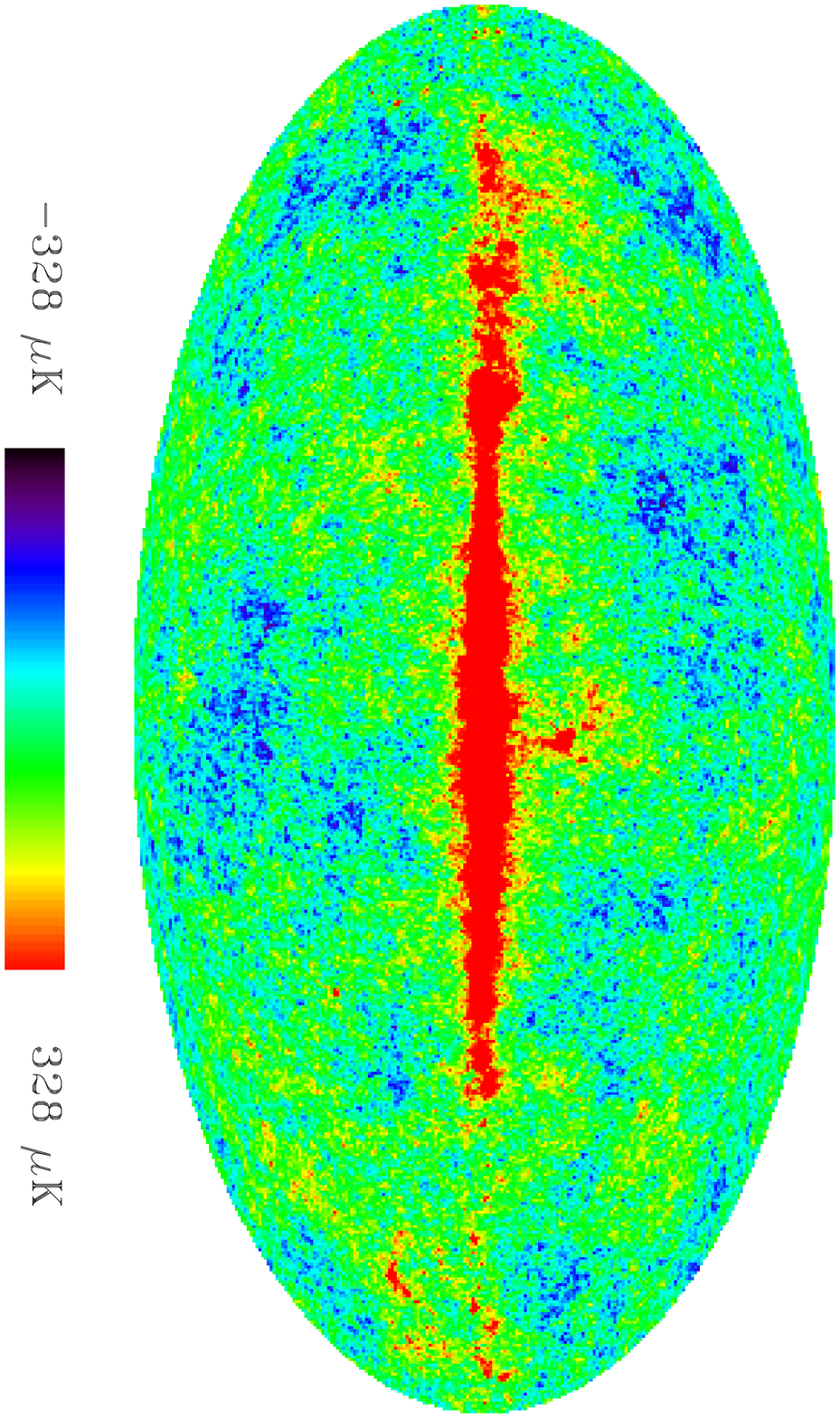}
}\caption[]{Simulated allsky reconstructed map at 143 GHz for
the 12 months of the Planck missions (average of 6 bolometers).
\label{fig3}} 
\end{figure}

\begin{figure}[ht]
\vbox to 20cm{
\epsfxsize=13.5cm
\epsfbox{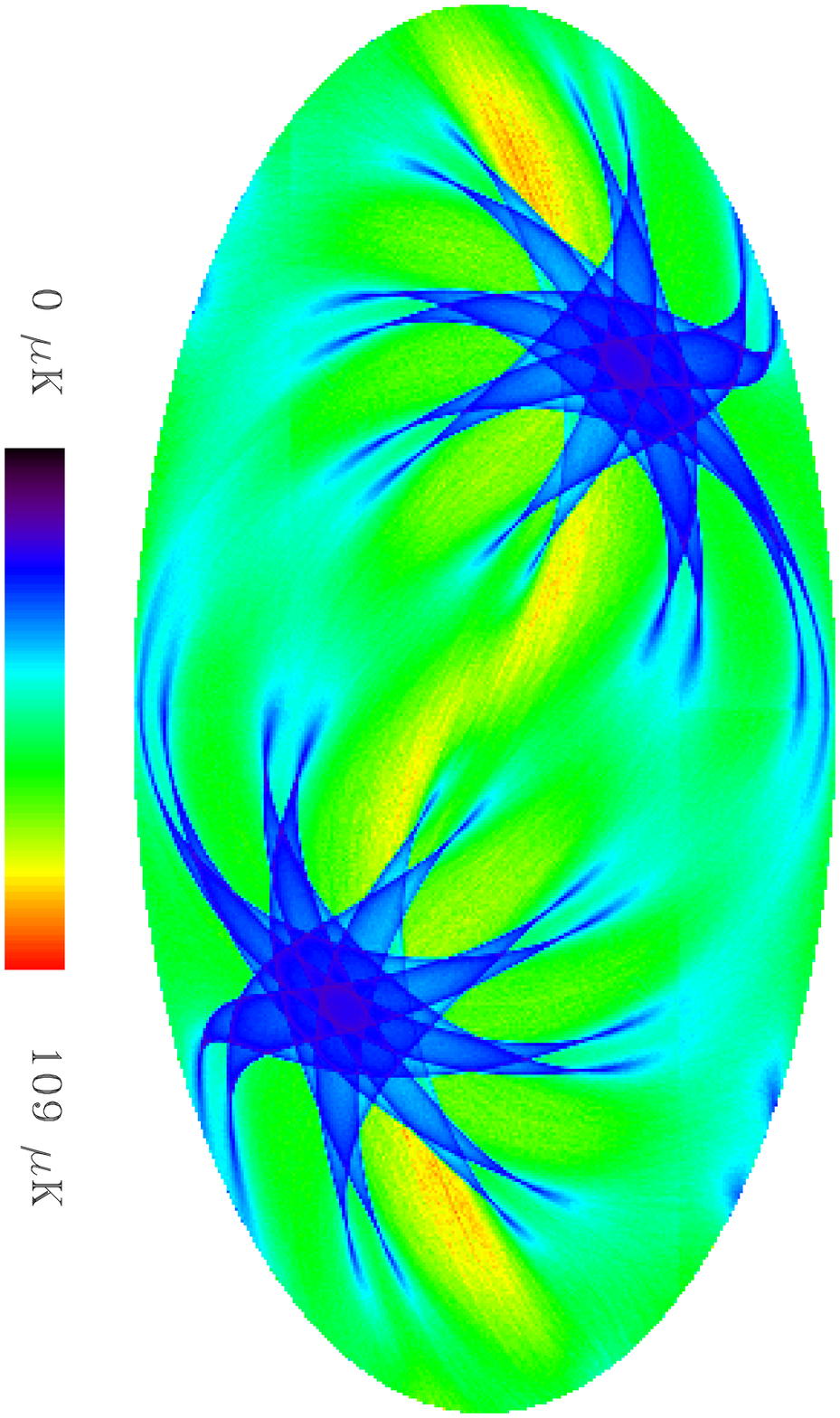}
}\caption[]{Variance map at 143 GHz
for 12 months of the Planck mission: no destriping
included.
\label{fig4a}} 
\end{figure}

\begin{figure}[ht]
\vbox to 20cm{
\epsfxsize=13.5cm
\epsfbox{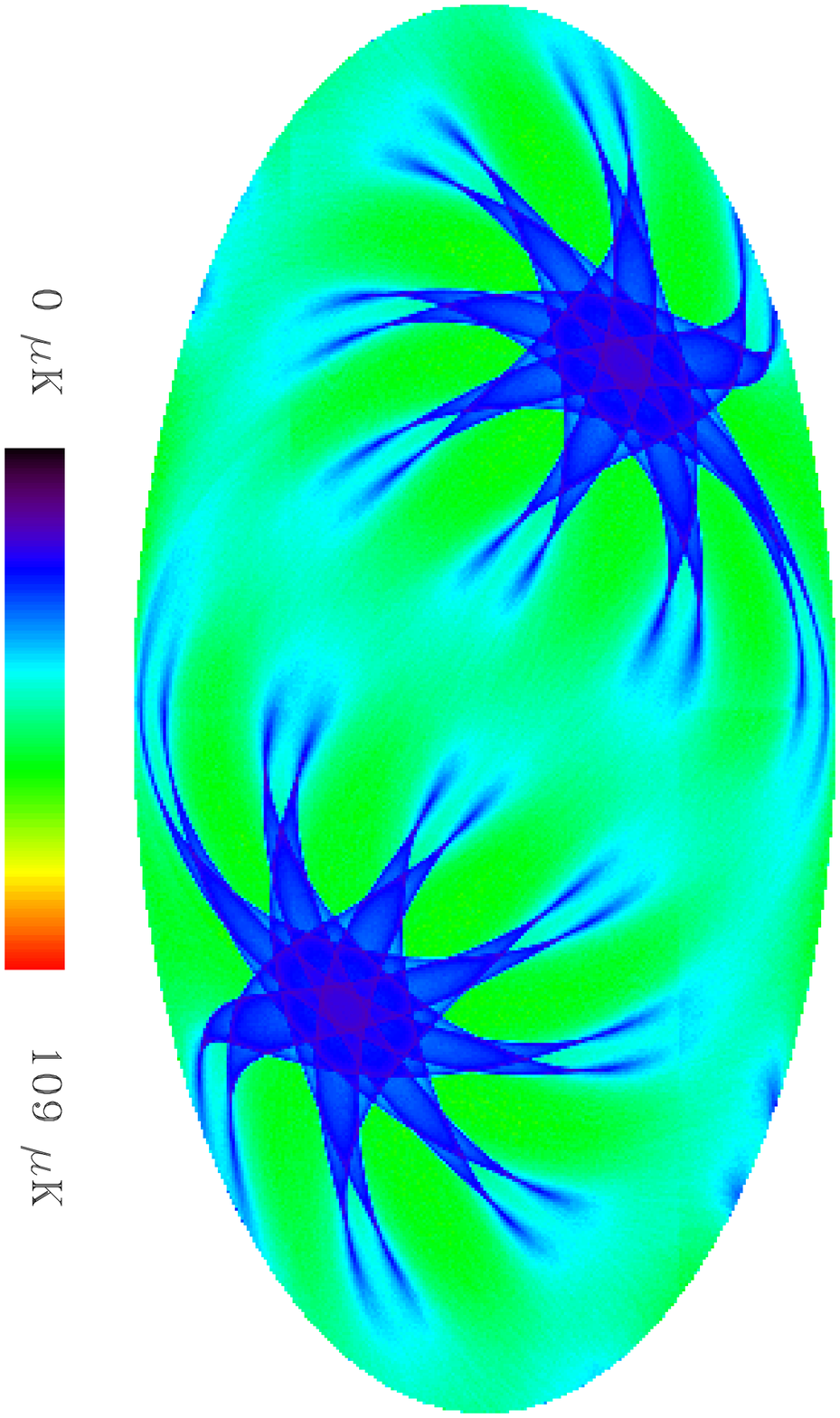}
}\caption[]{Variance map at 143 GHz
for 12 months of the Planck mission: destriping
included.
\label{fig4b}} 
\end{figure}

\begin{figure}[ht]
\vbox to 20cm{
\epsfxsize=13.5cm
\epsfbox{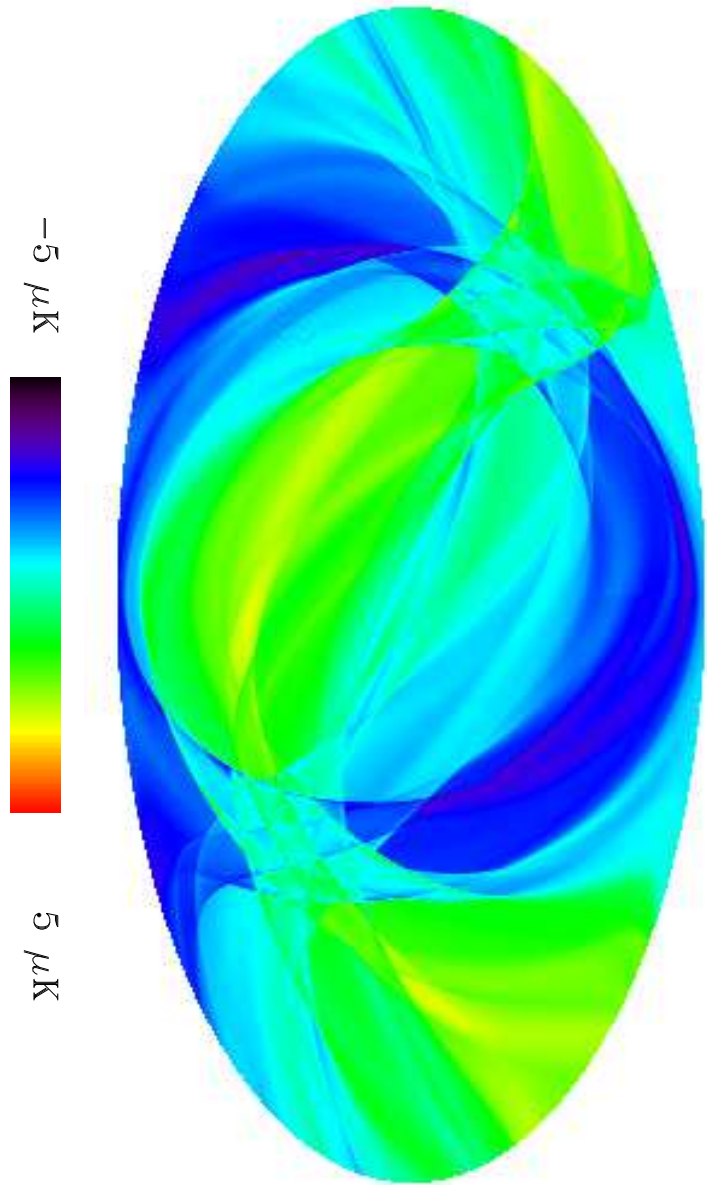}
}\caption[]{Map of the residual striping after 6 iterations of the
maximum gradient algorithm. This is the difference between the
input sky and the sky recovered after a simulation
with no instrumental noise. 
\label{fig4c}} 
\end{figure}

\begin{figure}[ht]
\vbox to 19cm{
\epsfxsize=13.5cm
\epsfbox{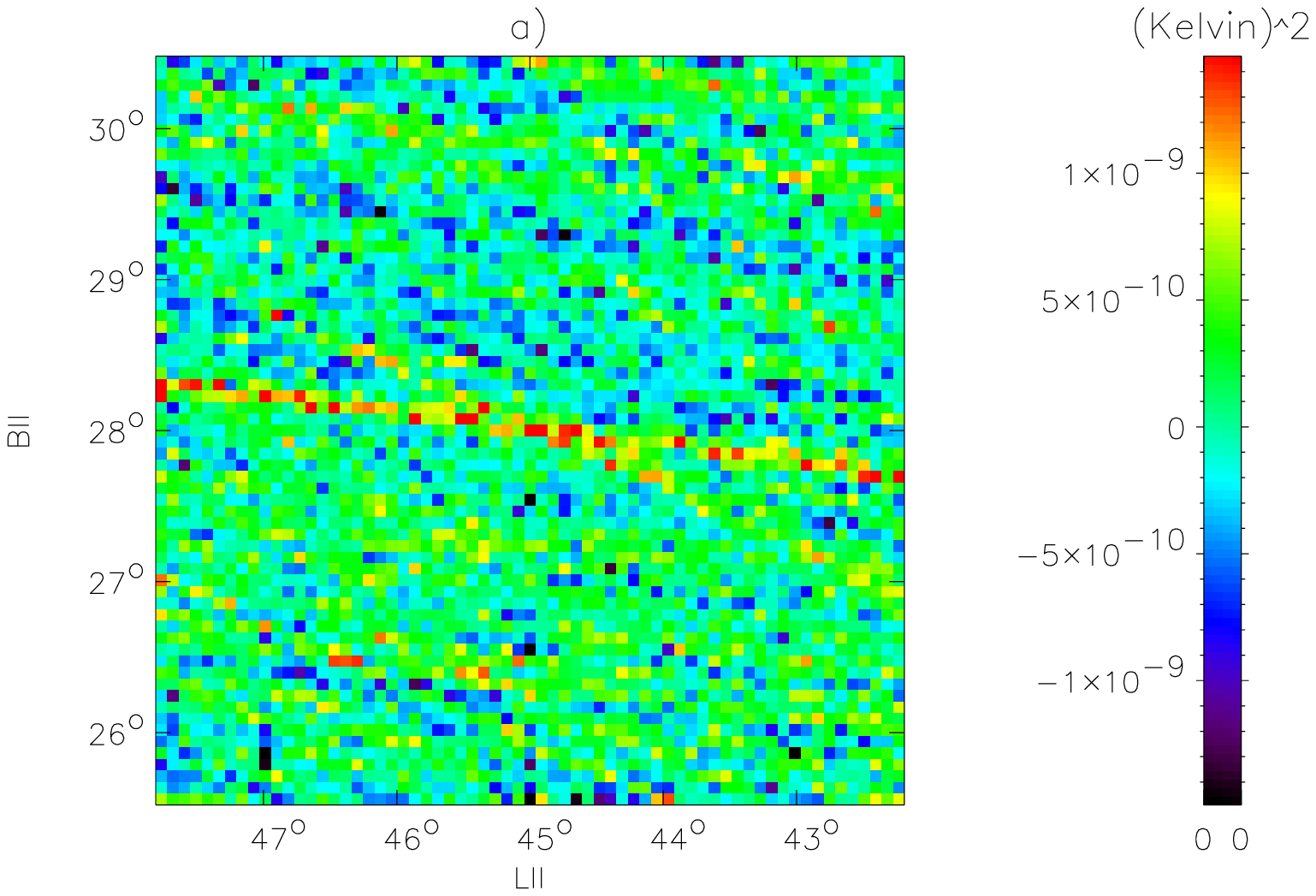}
\epsfxsize=13.5cm
\epsfbox{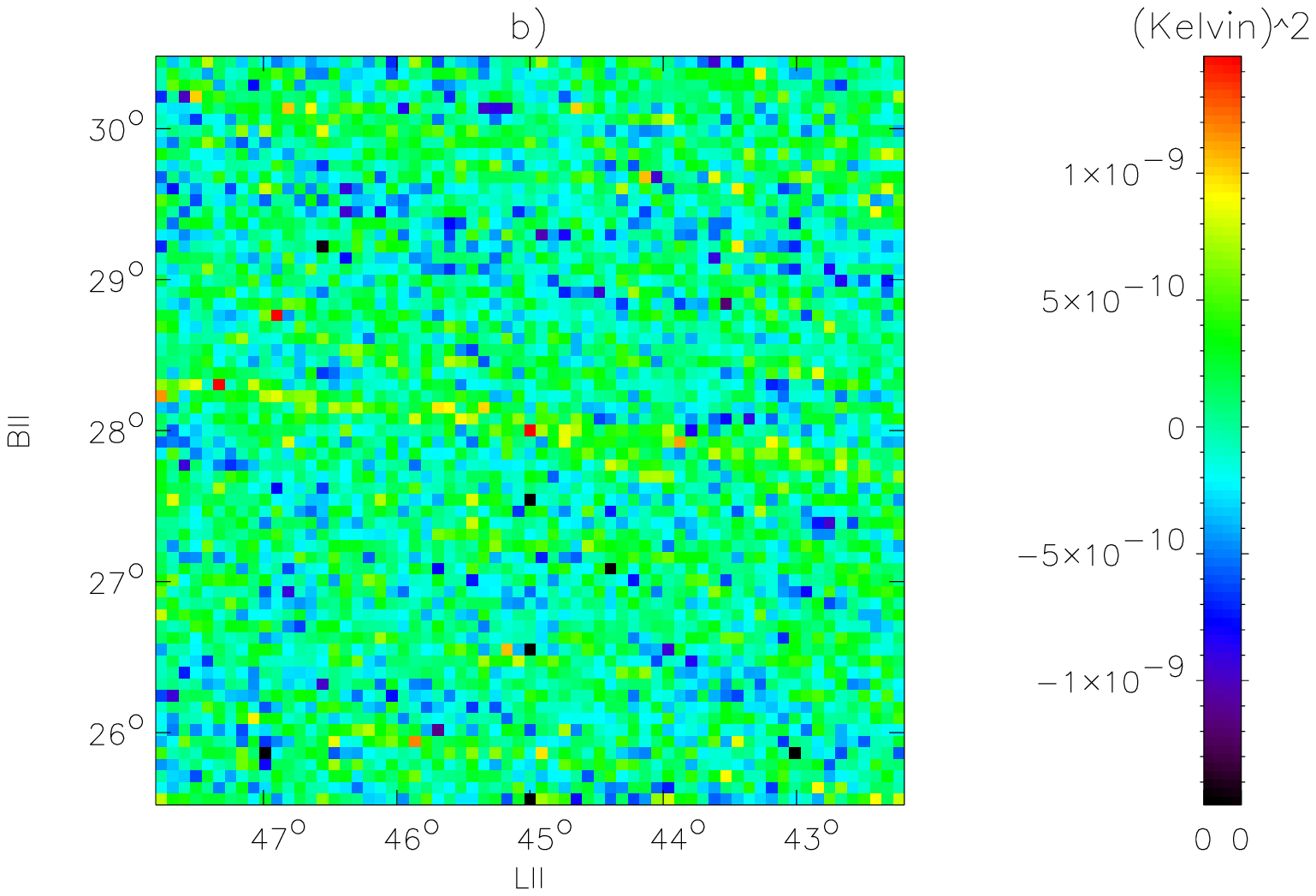}
}\caption[]{Estimated covariance map for a pixel at
$(LII, BII) = (45\deg, 28\deg)$ with its neighbour pixels : a) no destriping included, b) destriping
included.
\label{fig10}} 
\end{figure}

\end{document}